\documentclass[tradiabstract]{aa}  
\usepackage{graphicx}
\usepackage{txfonts}
\usepackage{natbib}

\bibpunct{(}{)}{;}{a}{}{,} 
\begin{document}

   \title{Injection to the pick-up ion regime from high energies and induced ion power laws}

   \author{H.-J. Fahr \inst{1}
   	\and
   	I. V. Chashei \inst{2}
          \and
          D. Verscharen \inst{1,3}}

\titlerunning{Injection to PUI regime and induced ion power laws}
\authorrunning{Fahr, H.-J. \& Chashei, I. V. \& Verscharen, D.}
   \institute{Argelander Institute for Astronomy, University of Bonn, Auf dem H\"ugel 71, 53121 Bonn, Germany
   	\and
   	Lebedev Physical Institute, Leninskii pr. 53, 117924 Moscow (Russia)
	\and
	Max Planck Institute for Solar System Research, Max-Planck-Str. 2, 37191 Katlenburg- Lindau, Germany
	}

   \date{}

 \abstract{
 Though pick-up ions (PUIs) are a well known phenomenon in the inner heliosphere,
 their phase-space distribution nevertheless is a theoretically unsettled
 problem. Especially the question of how pick-up ions form their suprathermal
 tails, extending to far above their injection energies, still now is
 unsatistactorily answered.
 Though Fermi-2 velocity diffusion theories have
 revealed that such tails are populated, they nevertheless show that
 resulting population densities are much less than seen in observations
 showing power-laws with a velocity index of ``-5''.
 We first investigate here, whether or not observationally suggested power-laws can be the result
 of a quasi-equilibrium state between suprathermal ions and
 magnetohydrodynamic turbulences in energy exchange with eachother. We
 demonstrate that such an equilibrium cannot be established, since it would
 require too high pick-up ion pressures enforcing a shock-free deceleration
 of the solar wind. We furthermore show that Fermi-2 type energy diffusion in
 the outer heliosphere is too inefficient to determine the shape of the
 distribution function there. As we can show, however, power-laws beyond the
 injection threshold can be established, if the injection takes place at
 higher energies of the order of 100 keV. As we demonstrate here, such an
 injection is connected with modulated anomalous cosmic ray (ACR) particles at the
 lower end of their spectrum when they again start being convected outwards
 with the solar wind. Therefore, we refer to these particles as ACR-PUIs. In our quantitative calculation of the pick-up ion
 spectrum resulting under such conditions we in fact find again power-laws,
 however with a velocity power index of ``-4'' and fairly
 distance-independent spectral intensities. As it seems these facts are
 observationally well supported by VOYAGER measurements in the lowest
 energy channels.

\keywords{Plasmas --
                solar wind --
                cosmic rays 
               }
}

\maketitle

\section{Introduction}

Suprathermal ions, picked-up by the supersonic solar wind flow as ionized
neutral atoms, have become known as pick-up ions (PUIs) and are produced
all over the inner heliosphere with a typical upwind-downwind asymmetry with respect to the inflow direction of the neutral ISM inflow vector \citep{rucinski93,fahr99}. In the
case of PUI protons, their production is due to photoionization and charge
exchange of interstellar H-atoms \citep[see][]{rucinski91,fahr99,rucinski03,bzowski07}. Their spatial
distribution seems well understood, while the PUI phase-space transport is a
much less settled subject. Especially it exists an ongoing debate of how
efficiently pick-up ions just after the pick-up process are accelerated to
higher energies due to nonlinear wave-particle interactions \citep[see e.g.][]{isenberg87,bogdan91,fichtner96,fichtner01,chalov96,chalov98,chalov04} and whether at all energy diffusion plays a relevant role in this
transport.

Some hint is given by the solar wind proton temperature behavior with
distance. The observed non-adiabatic temperature behavior namely proves that
a specific solar wind proton heating must operate in the outer heliosphere
which can only be due to energy absorption from pick-up ion generated
turbulence, since convected turbulence amplitudes quickly die out with
distance \citep[see][]{smith01,chashei02}.

Freshly injected PUIs represent keV-energetic protons in the supersonic
solar wind frame and may be called here: ``primary pick-up ions'' (or: PUIs$^{\ast }$). The velocity distribution of these newly produced PUIs$^{\ast }$
is toroidal and unstable \citep[see][]{winske84,winske85,lee87,fahr88}. With the free energy of this
unstable distribution PUIs$^{\ast }$ drive Alfv\'{e}nic wave power. The latter enforces
pitchangle isotropization of the initial velocity distribution \citep[see][]{chalov98,chalov99}. Due to wave-wave coupling, the wave energy generated
by PUIs$^{\ast }$ at the injection wavelength $\lambda _{i}=U_{s}/\Omega
_{p} $ is diffusively transported in wavevector space both to smaller
wavelengths where it can be absorbed by solar wind protons and to larger
wavelengths where it is reabsorbed by all PUIs. This effect is seen as the
main reason of solar wind proton heating occuring in the outer heliosphere
\citep{smith01,fahr02,chashei03,stawicki04}. Only a
small fraction of about 5 percent of the PUI-generated wave energy reappears
in the observed proton temperatures. Freshly injected PUIs excite turbulences that can organize a power-law distribution. From this distribution, both the  solar wind ions and the PUIs themselves can absorb energy as shown by \citet{chashei03}. Also the approach by \citet{isenberg03} where energy diffusion of pick-up ions is not taken into account shows that only a low degree (2-5 percent) of the pick-up ion driven wave energy is absorbed by solar wind protons in form of thermal energy.  This raises the question where the
major portion of the wave energy produced during the primary pick-up process
goes to. To clarify the energy redistributions, kinetic and spectral
details of the relevant processes have to be investigated. A detailed
numerical study of the PUI velocity distribution and the spectral
Alfv\'{e}nic/Magnetosonic wave power evolution has meanwhile been carried out
\citep{chalov04,chalov06a} and presents a simultaneous solution of a
coupled system of equations consistently describing the isotropic velocity
distribution function of PUIs and the spectral wave power intensity.

As one can see from this study, the largest portion of the self-generated
wave energy is reabsorbed by PUIs themselves as a result of the cyclotron
resonant interaction and leads to PUI-acceleration. It could perhaps be
hoped that this energization of pick-up protons due to Fermi-2 stochastic
acceleration processes eventually leads to the ubiquitous power-law PUI-tails
pointed out by \citet{fisk06,fisk07}. To the opposite, however, as
reflected in the results presented by \citet{chalov04,chalov06a} it is
evident that this is not the case: even at larger distances close to the
termination shock (100 AU) the PUI distributions show a rapid cut-off at
energies higher than the injection energy. The question thus is raised here
why power-laws have been seen at all. An explanation that we are favoring
here is a new injection source to the PUI regime from high energies
connected with modulated anomalous cosmic ray particles. These protons are primary ACR particles that occur with a spectrum down to the typical energy of the usually assumed PUIs. At this part of the spectrum, both particle species cannot be distinguished. Therefore, we refer to them as ACR-PUIs.

In Sect.~\ref{sect2} we investigate the physical possibility of power-law ions in the outer heliosphere as they are recently proposed by several authors and we find that they cannot occur with a power-index of $-5$. As we show in Sect.~\ref{sect3}, the proposed processes are not effective enough to produce the desired ion tails which means that another mechanism has to lead to the observed spectrum. In Sect.~\ref{sect4}, we show how a high-energy source can be derived by taking a modulated ACR spectrum upstream of the solar wind termination shock. This injection mechanism is discussed in Sect.~\ref{sect5} where we show that these high-energy ions can lead to power-law ion tails, however with a power-index of $-4$. The results are discussed and compared with observations in Sect.~\ref{sect6}

\section{Can power-law ion distributions be in equilibrium with hydromagnetic turbulence? }\label{sect2}

Challenged by recent results concerning ion spectra at large distances
\citep{mcdonald03,decker05,kiraly05,fisk06}, we look into the problem of PUI phase\-space transport under
these new given auspices. First we discuss the argument given by \citet{fisk06,fisk07} that pick-up ions under resonant interaction with
ambient compressive turbulences enter a quasi-equilibrium state with
saturated power-law distributions of a somehow sacrosanct spectral velocity index of
$\gamma _{v}=-5$.

The related, well known Kolmogorov formalism is based on a ``dimensional''
reasoning: The problem concerning the energy distribution in eddies of a
typical wave number $k$ is considered using two different dimensional
quantities: namely the spectral energy density $E_{k}$ and the wave number $k$ .
The spectral energy flux is then defined by
\begin{equation}\label{energy_flux}
S=(kE_{k})/\tau _{k}
\end{equation}
and has to be constant for stationary cases. $\tau _{k}$ is the typical life
time of eddies with scale $\lambda _{k}=2\pi /k$ and is given by 
\begin{equation}\label{tauk}
\tau _{k}=(kv_{k})^{-1}
\end{equation}
with the typical vortex velocity $v_{k}$ given by 
\begin{equation}
v_{k}=(E_{k}k)^{1/2}
\end{equation}

Requiring a constant energy flux $S$ given by Eq.(\ref{energy_flux}), with $\tau _{k}$
given by Eq.(\ref{tauk}) then automatically leads to the Kolmogorov spectrum in
the form $E_{k}$ $\sim k^{-5/3}$.

In their thermodynamical approach to particle spectra in equilibrium with waves, \citet{fisk06,fisk07}, however, only consider one
single dimensional quantity, i.e.~the kinetic energy $T$, because in their
case the energy distribution $\digamma$ has the dimension $\left[ \digamma %
\right] =\left[ T\right] ^{-1}$. The flux combination
\begin{equation}
S_{F}=T\cdot \digamma \cdot \frac{\Delta T}{\Delta t}
\end{equation}
is then expected by them to be constant and, thus, the energy gain is given by
\begin{equation}
\frac{\Delta T}{\Delta t}\sim T
\end{equation}
and hence the above requirements result in the requirement that $\digamma
\sim T^{-2}$.

This shows that the analogy of the Fisk-Gloeckler approach with the
Kolmogorov formalism is not complete, since in their theory the quantity $\Delta t$ is not specified. Only if we play the same game with the Kolmogorov
turbulence using instead of Eq.(\ref{tauk}) the assumption $\tau =(ku)^{-1}$,
where $u$ is some independent external speed (especially independent on $k$),
we will find from Eq.(\ref{energy_flux}) also the result
\begin{equation}
E_{k} \sim k^{-2}
\end{equation}
analogous to the result obtained by Fisk-Gloeckler.

The problem may be briefly inspected here whether or not some
spatial/temporal disturbances in the solar wind plasma can be considered as
waves. They should be considered as waves, if the convection time $\tau
_{\mathrm{conv}}=r/U$ is much greater than the passage period of these waves in the
wind reference frame $\tau _{\sim }=2\pi /(k_{0} v_{\mathrm A})$ where $r$ denotes the radial distance to the sun and $U$ the solar wind bulk velocity. The Alfv\'en speed is designated as $v_{\mathrm A}$ with the corresponding wavevector $k_0$ for the turbulence. For an
inequality of the kind $\tau _{\sim }\ll t_{\mathrm{conv}}$, this then leads to the
definition of the principal wave turbulence correlation scale given by $%
L_{\mathrm m}<rv_{\mathrm A}/U\approx 0.1\,\mathrm{AU}$. A more exact definition of the
turbulence correlation scale is given in the paper of \citet{chashei03}.
This value is also in accordance with the value used by \citet{fahr07b} for his
estimation of the upper possible velocity border. Some measurements indicate smaller values for the turbulence correlation length  than calculated by us in this paper \citep[see e.g.][]{matthaeus05}. However, the following conclusions in our paper drawn on the effects of  turbulent heating of suprathermal protons become even less promising for an MHD-equilibrium to be established,  if smaller correlation lengths prevail. One can expect that at
larger distances, $r\geq 1\,\mathrm{AU}$, the value for $L_{\mathrm m}$ increases proportionally
to $r$, since a similar dependence is also valid for the outer scale of turbulence $k_{0}^{-1}$ \citep[see][]{chashei03}. Here we estimate the maximum
energy for protons resonating with these largest scales $L_{\mathrm m}$ at distances
of about 100 AU and find with $v_{\max }\simeq \Omega (100\,\mathrm{AU})\cdot
L_{\mathrm m}(100\,\mathrm{AU})/\gamma (v_{\max })\simeq \Omega (1\,\mathrm{AU})\cdot L_{\mathrm m}(1\,\mathrm{AU})/\gamma
(v_{\max })$, where $v_{\max }$ , $\gamma (v_{\max })$ and $\Omega $ are the
maximal ion speed, the associated Lorentz factor and the ion gyrofrequency,
respectively. Protons with these speeds have energies of about $E_{\max
}\approx 1\,\mathrm{GeV}$ independent on distance. Thus, ions with $E\leq E_{\max }$ can
be expected to be scattered by waves, whereas ions with $E\geq E_{\max }$
can only be scattered by large scale velocity structures in the solar
wind.

Since in any case, however, power-laws seem to be a fact well supported by
observations within a certain range of solar distances and ion energies, we
shall nevertheless take this finding as serious here as deserved and
determine in the following the absolute spectral intensity of this PUI
power-law distribution and its consequences.

If the PUI-distribution is given in the form
\begin{equation}
f_{\mathrm{pui}}(r,v)=f_{\mathrm{pui},0}\cdot \left(\frac{v}{v_{0}}\right)^{\gamma _{v}}
\end{equation}
with the velocity power index $\gamma _{v}=-5$, then the PUI density is
given by
\begin{equation}
n_{\mathrm{pui}}(r)=4\pi f_{\mathrm{pui},0}\cdot \int \limits_{v_{0}}^{v_{\infty
}}\left(\frac{v}{v_{0}}\right)^{-5}v^{2}\mathrm dv\simeq 2\pi f_{\mathrm{pui},0}\cdot v_{0}^{3}
\end{equation}
where $f_{\mathrm{pui},0}=f_{\mathrm{pui},0}(r)$ is a local normalization value, and $v_{0}$
and $v_{\infty }$ are lower and upper velocity limits of the quasistationary
PUI power law.

Using $\psi =v/v_{0}$, one obtains for the power-law PUI pressure as the second moment of the distribution function the
following result
\begin{equation}\label{puipress}
P_{\mathrm{pui}}(r)=4\pi f_{\mathrm{pui},0}\frac{m}{2}v_{0}^{5}\int\limits_{1}^{\psi _{\infty
}} \psi^{\prime -5}\psi^{\prime 4}\mathrm d\psi^{\prime}=2\pi m\cdot f_{\mathrm{pui},0}v_{0}^{5}\ln (\psi _{\infty })
\end{equation}

Evidently, the definition of $P_{\mathrm{pui}}(r)$ requires the determination of all
three local values $f_{\mathrm{pui},0}(r)$, $v_{0}(r)$, and $v_{\infty }(r)$ which
we aim at below.

\subsection{Determination of the inner and outer velocity border}

Using the definition of the lower velocity border as done by
\citet{fahr07b} one finds
\begin{equation}\label{nunull}
v_{0}\simeq \frac{1}{5}U\left( \frac{r}{r_{\mathrm E}}\right) ^{1/4}.
\end{equation}

With the above result one obtains the PUI distribution in the form
\begin{equation}\label{fpuinull}
f_{\mathrm{pui},0}=n_{\mathrm{pui}}(r)\cdot \frac{1}{2\pi v_{0}^{3}}=\frac{5^{3}}{2\pi U^{3}}%
n_{\mathrm{pui}}(r)\cdot \left( \frac{r}{r_{\mathrm E}}\right) ^{-3/4}
\end{equation}

The local PUI density can be derived from the interplanetary H-atom density $%
n_{\mathrm H}(r,\theta )$, which depends on the radial distance $r$ and the inclination angle $\theta$ with respect to the upwind axis, and the effective (charge exchange + photoionization)-
induced injection rate $\beta _{\mathrm{pui}}=n_{\mathrm H}(r,\theta )\left[n_{\mathrm s}(r)\sigma
_{\mathrm{ex}}U+\nu _{\mathrm i}\right]$. Here $n_{\mathrm s}(r)$, $\sigma _{\mathrm{ex}}$, $U$, $\nu _{\mathrm i}$ denote
the solar wind proton density, the charge exchange cross section, the solar
wind bulk velocity and the photoionization frequency. The H-atom density at
distances $r\geq 5\,\mathrm{AU}$ in the upwind hemisphere is satisfactorily well given by the following
expression \citep[see][]{fahr71}
\begin{equation}
n_{\mathrm H}(r,\theta )=n_{\mathrm H,\infty }\exp \left[-\frac{\beta _{\mathrm{pui},0}r_{0}^{2}\theta }{%
Ur\sin \theta }\right]
\end{equation}
and thus leads to the following PUI density \citep[also see][]{fahr99}
\begin{equation}\label{npui}
n_{\mathrm{pui}}(r,\theta )=\left(\frac{r_{0}}{r}\right)^{2} n_{\mathrm{pui},0}+\frac{1}{r^{2}U}%
\int \limits_{r_{0}}^{r}\beta _{\mathrm{pui}}(r^{\prime},\theta )r^{\prime 2}\mathrm dr^{\prime}
\end{equation}

From the above one derives the following radial space derivative which
lateron in this paper will be needed
\begin{equation}\label{ngrad}
\frac{\partial n_{\mathrm{pui}}(r,\theta )}{\partial r}=\frac{-2n_{\mathrm{pui}}(r,\theta )%
}{r}+\frac{\beta _{\mathrm{pui}}(r,\theta )}{U}
\end{equation}

To calculate the PUI pressure, one more quantity in addition is
needed, namely the upper velocity border $v_{\infty }$. To determine $%
v_{\infty }$ we follow the idea presented by \citet{fisk06,fisk07}
assuming that PUI power law distributions result from a specific
quasi-equilibrium state self-establishing such that the wave field transfers
per unit of time as much energy to PUIs by energy diffusion, as energy is
expended in the solar wind frame for the work done by the pressure gradient
of comoving PUIs against the magnetosonic fluctuations. The important restriction to energy diffusion by nonlinear interaction with the compressive fluctuations is that the typical diffusion period $\tau_{\mathrm{diff}}\simeq 3L_{\mathrm m}^2/v\lambda_{\parallel}$ should be much larger than the convection period given by $\tau_{\mathrm{conv}}\simeq L_{\mathrm m}/U$ \citep[see][]{chalov03}. This leads to the requirement
\begin{equation}
L_{\mathrm m} > \frac{v\lambda_{\parallel}}{3U}
\end{equation}
with $\lambda_{\parallel}$ as the mean free path for particles parallel to the magnetic field. The uppermost velocity $v_{\infty}$ is the limit at which this condition is just fulfilled:
\begin{equation}
v_{\infty}\simeq \frac{3UL_{\mathrm m}}{\lambda_{\parallel}}.
\end{equation}
We assume the interaction of the particles with a slab Alf\`{e}nic turbulence field. The mean free path is given by 
\begin{equation}
\lambda_{\parallel}=\frac{3v}{8} \int \limits _{-1}^{+1} \frac{(1-\mu^2)^2}{D_{\mu \mu}} \mathrm d\mu
\end{equation}
with the pitch-angle diffusion coefficient
\begin{equation}
D_{\mu \mu}=D_{vv,0} v_{\mathrm A}^{-2} \left(\frac{U^3}{r_{\mathrm E}} \right)\left(\frac{v}{U} \right)\left(\frac{r_{\mathrm E}}{r} \right)^{3/4}
\end{equation}
from \citet{chalov03} for cyclotron resonant wave-particle interaction with unpolarized, one-dimensional, and isotropic turbulence, which leads to the velocity independent expression
\begin{equation}
\lambda_{\parallel}=\frac{2}{5} v_{\mathrm A}^2 \left(\frac{r_{\mathrm E}}{U^3} \right) \left(\frac{U}{D_{vv,0}}\right) \left(\frac{r_{\mathrm E}}{r} \right)^{3/4}=\lambda_{\parallel,\mathrm{E}}\left(\frac{r_{\mathrm E}}{r} \right)^{3/4}
\end{equation}
for the mean free path, with the reference value
\begin{equation}
D_{vv,0}=\frac{\sqrt{\langle\delta u^2 \rangle_{\mathrm E}}}{U}\frac{r_{\mathrm E}}{9L_{\mathrm m}}
\end{equation}
for the diffusion coefficient. Therefore, the upper velocity border is given by
\begin{equation}
v_{\infty}\simeq \frac{3UL_{\mathrm m}}{\lambda_{\parallel,\mathrm E}},
\end{equation}
which evaluates to
\begin{equation}
v_{\infty}\simeq \frac{9}{0.3}\left(\frac{r_{\mathrm E}}{r} \right)^{3/4} U
\end{equation}
by taking $L_{\mathrm m}=3\,\mathrm{AU}$ and $\lambda_{\parallel,\mathrm E}=0.3 \,\mathrm{AU}$ \citep{chalov_fahr99}.
The ratio
\begin{equation}\label{psi}
\psi_{\infty}=\frac{v_{\infty}}{v_0}=150x^{-1}
\end{equation}
is needed in the later calculation and should be compared with the result obtained by \citet{fahr07b} deriving the upper velocity border from the study of the upper-most
resonance possibilities of ions with the largest prevailing correlation
lengths $L_{\mathrm m}$ existing in the solar wind velocity structures, yielding the
result
\begin{equation}
\psi _{\infty }=30\cdot x^{-3/4}
\end{equation}
which gives smaller values than those derived for conditions
when balanced pressure equilibrium in the sense of \citet{fisk07}
is adopted. Here below we shall demonstrate that this value for $\psi _{\infty }$, which is the direct consequence of the assumptions by Fisk, definitely leads to unreasonable consequences when we
investigate the associated PUI pressure.

\subsection{The upstream pick-up ion pressure }

We calculate the PUI pressure resulting from power-law distributed PUIs
upstream of the shock and find with Eqs.~(\ref{puipress}), (\ref{nunull}), (\ref{fpuinull}), and (\ref{psi})
\begin{equation}
P_{\mathrm{pui}}(x)=mU^2n_{\mathrm{pui}}(x)x^{1/2} \ln\left(150x^{-1}\right),
\end{equation}
where we have introduced $x=r/r_{\mathrm E}$.
We obtain the pressure gradient by differentiation:
\begin{eqnarray}
&&\frac{\partial P_{\mathrm{pui}}}{\partial x}=m U^2 \times \\
&\times &\left[ \frac{\partial n_{\mathrm{pui}}}{\partial x}x^{1/2}  \ln\left(150x^{-1}\right)+ \frac{1}{2}n_{\mathrm{pui}}x^{-1/2} \ln\left(150x^{-1}\right) - n_{\mathrm{pui}}x^{-1/2}  \right] \nonumber
\end{eqnarray}
which leads with Eq.~(\ref{ngrad}) to
\begin{eqnarray}
&&\frac{\partial P_{\mathrm{pui}}}{\partial x}=m U^2 \times  \\
&\times& \left[ -\frac{3}{2}n_{\mathrm{pui}}x^{-1/2}\ln\left(150x^{-1}\right) +\frac{\beta r_{\mathrm E}}{U}x^{1/2}\ln\left(150x^{-1}\right)-n_{\mathrm{pui}}x^{-1/2}\right].\nonumber
\end{eqnarray}

If now we derive the effective upstream Mach number, neglecting thereby
solar wind electron and proton pressures compared to the PUI pressure, i.e
assuming $P_{e};P_{s}\ll P_{\mathrm{pui}}$, and following the definitions by \citet{fahr99},
we obtain
\begin{eqnarray}
M_{\mathrm s}^{\ast 2}&=&\left(\frac{U}{C_{s}^{\ast }}\right)^{2}=\frac{U^{2}\frac{\partial \rho
_{\mathrm s}}{\partial r}}{\frac{\partial }{\partial r}(P_{\mathrm s}+P_{\mathrm e}+P_{\mathrm{pui}})}\simeq\\
&\simeq& \frac{2n_{\mathrm s}}   {\frac{3}{2}n_{\mathrm{pui}}x^{1/2}\ln\left(150x^{-1}\right) -\frac{\beta r_{\mathrm E}}{U}x^{3/2}\ln\left(150x^{-1}\right)+n_{\mathrm{pui}}x^{1/2}}.\nonumber 
\end{eqnarray}
With $\Lambda=\frac{\beta r_0}{U n_{\mathrm s}}\simeq 10^{-2}$ and the PUI abundance $\xi=n_{\mathrm{pui}}/n_{\mathrm s}\simeq 0.2$ \citep[see][]{fahr99}, we obtain
\begin{equation}
M_{\mathrm s}^{\ast 2}=\frac{2x^{-1/2}}{\xi \left(\frac{3}{2}\ln\left(150x^{-1}\right)+1 \right) - \Lambda \frac{r_{\mathrm E}}{r_0}x\ln\left(150x^{-1}\right)}.
\end{equation}
For a shock position at $x\simeq 100$ and with the lower reference distance $r_0\simeq 5\,\mathrm{AU}$, this expression yields
\begin{equation}
M_{\mathrm s}^{\ast 2}=0.83,
\end{equation}
which means that the effective upstream Mach number $M_{\mathrm s}^{\ast}=0.91$ is lower than 1.

Hence, power-law PUI pressures in balance with the wave fields would not
allow for the occurence of a termination shock of the solar wind (i.e., needing upstream Mach numbers $M_{\mathrm s}^{\ast }\geq 1$!)

The non-existence of  the above ``turbulence - particle'' - equilibrium
(TPE) can also be concluded along a different line of argumentations
together with the calculation of the downstream PUI pressure connected with
power-law distributed PUIs. Using the results presented in \citet{fahr00}, one obtains on the basis of Liouville's theorem and conservation of
magnetic moment at the passage from upstream to downstream over the shock
the PUI distribution function downstream of the shock (indicated by index 2) by the following
relation
\begin{equation}
f_{\mathrm{pui},2}(v)=\frac{1}{\sqrt{s}}f_{\mathrm{pui},1}\left(\frac{v}{\sqrt{s}}\right)
\end{equation}
depending on the upstream distribution function $f_{\mathrm{pui},1}$,
where $s$ denotes the compression ratio at the shock. From that relation one
obtains for the downstream PUI pressure $P_{\mathrm{pui},2}$ the following expression
\begin{eqnarray}
P_{\mathrm{pui},2}&=&\frac{m}{2\sqrt{s}}\int\limits_{\sqrt{s}v_{0}}^{\sqrt{s}v_{\infty
}}f_{\mathrm{pui},1}(v/\sqrt{s})v^{4}\mathrm dv \nonumber \\
&=&\frac{m}{2} s^{2}\int \limits_{v_{0}}^{v_{\infty}} f_{\mathrm{pui},1}(v)v^{4}\mathrm dv=s^{2} P_{\mathrm{pui},1}
\end{eqnarray}
which states that the downstream PUI pressure is enhanced with respect to
the upstream PUI pressure by the factor $s^{2}$. Reminding that the latter
pressure for TPE-conditions is given by
\begin{equation}
P_{\mathrm{pui},1}(r_{\mathrm{sh}})=mn_{\mathrm{pui}}(r_{\mathrm{sh}})U^{2}\left(\frac{r_{\mathrm{sh}}}{r_{\mathrm E}}\right)^{1/2}\ln\left(150\frac{r_{\mathrm E}}{r_{\mathrm{sh}}}\right)
\end{equation}
one finds that the downstream PUI pressure should amount to
\begin{equation}
P_{\mathrm{pui},2}(r_{\mathrm{sh}})=mn_{\mathrm{pui}}(r_{\mathrm{sh}})s^2U^{2}\left(\frac{r_{\mathrm{sh}}}{r_{\mathrm E}}\right)^{1/2}\ln\left(150\frac{r_{\mathrm E}}{r_{\mathrm{sh}}}\right)
\end{equation}
which normalized with the upstream solar wind kinetic energy density $\epsilon _{\mathrm{kin}}=(1/2)mU^{2}(n_{\mathrm s})$ would require that
\begin{equation}
 \frac{P_{\mathrm{pui},2}(r_{\mathrm{sh}})}{\epsilon _{kin}(r_{\mathrm{sh}})}=2\xi
(r_{\mathrm{sh}})s^{2}\left(\frac{r_{\mathrm{sh}}}{r_{\mathrm E}}\right)^{1/2} \ln\left(150\frac{r_{\mathrm E}}{r_{\mathrm{sh}}}\right)
 \end{equation}
with $\xi (r_{\mathrm{sh}})\simeq 0.2$ denoting the PUI abundance at the
shock. This would mean that the downstream thermal energy of the PUIs is
much higher than the kinetic energy of the upstream solar wind which is forbidden by
physical reasons. This again leaves to conclude that PUIs cannot exist in pressure equilibrium with the compressional magnetosonic turbulence that was assumed by \citet{fisk07} and cannot be responsible for the stochastic particle acceleration up to regions near the solar wind termination shock.

Requiring that the downstream thermal energy of the PUIs stays below the
upstream kinetic energy of the PUIs would require instead an upper border $%
v_{\infty }$ of the PUI power spectrum defined by
\begin{eqnarray}
P_{\mathrm{pui},2}&=&\frac{ms^{2}}{2}\int \limits_{v_{0}}^{v_{\infty }}f_{\mathrm{pui},1}(v)v^{4}\mathrm dv \nonumber \\
&=&s^{2}\left[4\pi f_{\mathrm{pui}}\frac{m}{2}v_{0}^{5}\ln (\psi_{\infty
})\right]\leq \frac{m}{2}U^{2}(n_{\mathrm s}+n_{\mathrm{pui}})
\end{eqnarray}
yielding
\begin{equation}
\ln (\psi_{\infty })\leq \frac{U^{2}(n_{\mathrm s}+n_{\mathrm{pui}})}{s^{2}4\pi
f_{\mathrm{pui}}v_{0}^{5}}=\frac{U^{2}(n_{\mathrm s}+n_{\mathrm{pui}})}{2s^{2}n_{\mathrm{pui}}v_{0}^{2}}=\frac{%
25}{2s^{2}\xi _{\mathrm{pui}}}\left(\frac{r_{\mathrm{sh}}}{r_{\mathrm E}}\right)^{-1/2}
\end{equation}

Evaluating this formula with $s=2.5$, $\xi _{\mathrm{pui}}=0.2$, and $%
r_{\mathrm{sh}}=100r_{\mathrm E}$ leads to the result
\begin{equation}
\ln (\psi_{\infty })\leq \frac{25}{2\cdot 2.5^{2}\cdot 0.2\cdot 10}=1
\end{equation}
and means that $\psi_{\infty }=\exp (1)=2.72$ and $v_{\infty }\simeq 2.72U$, i.e.~much lower than required for pressure equilibrium
conditions.

\section{The relative effectiveness of energy diffusion and convective
changes at ion phasespace transport}\label{sect3}

In the following we want to clarify the role of energy diffusion in
determining the shape of the PUI distribution function. We start from the
transport equation adequate to describe the phasespace behavior of the
PUIs by a distribution function $f(t,r,v)$ \citep[see e.g.][]{isenberg87,chalov96}
\begin{eqnarray}
\frac{\partial f}{\partial t}&+&\vec{U}\frac{\partial f}{\partial \vec{r}}-\left(\frac{v}{3}\right) \left( \frac{\partial f}{\partial v}\right) \mathrm{div}\vec{U}=\frac{1}{v^{2}}\frac{\partial}{\partial v}\left(v^{2}D_{vv}\frac{\partial f}{\partial v}\right)+ \nonumber \\
&+&Q(\vec{r,}v)+S(\vec{r},v) \label{transport}
\end{eqnarray}
where $\vec{U}$ is the solar wind bulk speed, $D_{vv}$ is the velocity
diffusion coefficient, and, $Q(\vec{r},v)$ and $S(\vec{r},v)$ are
functions describing PUI- injection sources and - phasespace losses. Terms
on the left hand side of the above Eq.~(\ref{transport}) under steady state conditions induce
changes with a typical convection time $\tau _{\mathrm{conv}}$ given by
\begin{equation}\label{tconv}
\tau _{\mathrm{conv}}\approx r/U
\end{equation}
where $r$ is of the order of the heliocentric distance.

Considering quasi-linear velocity diffusion of particles due to Fermi-2 type
interactions with the Alfv\'{e}nic- or magnetosonic turbulence, both of which
are leading to analogous expressions \citep[see][]{toptygin85,leroux98,chalov00}. For our estimates here,
one can use the diffusion coefficient derived by \citet{schlickeiser89} which
for estimate purposes can be represented in the following form
\begin{equation}\label{dvv}
D_{vv}\approx \delta _{\mathrm M0}\Omega v_{\mathrm A}^{2}\left(\frac{k_{0}}{k_{\mathrm{res}}}\right)^{\alpha -1}
\end{equation}
where $k_{\mathrm{res}}=\Omega /v$ is the resonant wave number, $\Omega $ the proton
cyclotron frequency, $v_{\mathrm A}$ the Alfv\'{e}n speed, $k_{0}$ the
turbulence outer scale, $\alpha \simeq 5/3$ (or $\simeq 3/2$)  the
power exponent of the 1D- turbulence spectrum, and $\delta _{\mathrm M0}$ is the
fractional turbulence level
\begin{equation}\label{deltamnull}
\delta _{\mathrm M0}=\frac{\langle\delta \vec{B}^{2}\rangle}{\langle\vec{B\rangle}^{2}}
\end{equation}
where $\vec{B}$ is the induction of local interplanetary magnetic field.
Typical velocity diffusion times $\tau _{\mathrm{diff}}$ characterizing the action
of the first term of the right-hand side of Eq.~(\ref{transport}) in changing the
distribution function $f$ can be defined by
\begin{equation}\label{tdiff}
\tau _{\mathrm{diff}}^{-1}\approx D_{vv}v^{-2}
\end{equation}
Combining the relations given by Eqs.(\ref{tconv}), (\ref{deltamnull}), (\ref{dvv}), and (\ref{tdiff}) yields as a
typical ratio of the characteristic times
\begin{equation}
\chi (v,r)=\frac{\tau _{\mathrm{conv}}}{\tau _{\mathrm{diff}}}\approx \delta
_{\mathrm M0}\left(\frac{v_{\mathrm A}^{2}}{vU}\right)(k_{\mathrm{res}}r)\left(\frac{k_{0}}{k_{\mathrm{res}}}\right)^{\alpha -1}\sim
v^{\alpha -3}
\end{equation}
At heliocentric distances $r\approx r_{\mathrm E}=1\,\mathrm{AU}$, one can assume \citep{chashei03} $\delta _{\mathrm M0}\approx 0.1$, $v_{\mathrm A}\approx
0.1U$, $k_{0}\approx 10^{-11}\mathrm{cm}^{-1}$, $\alpha =3/2$ for
Iroshnikov-Kraichnan turbulence (i.e., a power law for the spectral energy density $E_k\sim k^{-3/2}$) or $\alpha =5/3$ for Kolmogorov turbulence (i.e., $E_k\sim k^{-5/3}$), which are the two mostly found and discussed spectral energy distributions in solar wind turbulence \citep{bale05,horbury05}. Then near 1 AU, we find a ratio $\chi (U,1\,\mathrm{AU})\approx 1$
showing that convection and diffusion processes here are of comparable
importance for the particles with $v\approx U$, while convection
effects are of increasing importance for particles with higher velocities,
i.e.~with $v>U$. This result is in good agreement with other considerations   
\citep{matthaeus_goldstein86,jokipii_kota89,zank_matthaeus92,zank96,horbury_balogh01}.
\citet{isenberg05}, however, concludes that the effect of energy diffusion is small for PUIs. His similar treatment is parametrized by the percentage of turbulence occuring in slab Alfv\`{e}nic fluctuations. Nevertheless, these two approaches can be conciliated if this percentage is assumed to amount about 20\%.

At larger heliocentric distances, $r>1\,\mathrm{AU}$, the frozen-in Parker magnetic
field near the ecliptic is nearly azimuthal and decreases with $(1/r)$.
Correspondingly, the resonant wave vector behaves like $k_{\mathrm{res}}\sim
\Omega \sim \left\langle B\right\rangle \sim r^{-1}$,
which leads to $k_{\mathrm{res}}r=\mathrm{const}$. Besides that, as argued by \citet{chashei03}, the following $r$-dependences can be expected: $k_{0}\sim r^{-1}$
and $\delta _{\mathrm M0}\sim r^{-2/3}$. Consequently, the radial dependence of $\chi (v,r)$ is exclusively determined by the $r$-dependence of $\delta _{\mathrm M0}$
meaning that
\begin{equation}\label{chir}
\chi (v,r)\approx \left(\frac{r}{r_{\mathrm E}}\right)^{-2/3}\left(\frac{v}{U}\right)^{-7/3}
\end{equation}
in case of Kolmogorov turbulence. The above Eq.~(\ref{chir}) shows that at larger
solar distances energy diffusion of particles cannot determine the shape of
the suprathermal tail of the resulting PUI distribution function.

In fact, assuming for reasons of a better clarification, just to the
contrast here, a dominance of the energy diffusion, i.e.~neglecting all the
terms in Eq.~(\ref{transport}) except for the diffusion term, would deliver as a result of
the transport equation the distribution $f\sim v^{-\alpha }$. This
distribution thus should evidently be much flatter than the obviously
observed distribution $f\sim v^{-5}$ \citep{fisk06}
indicating that energy-diffusion in fact plays an inferior role.

This result also turned out from more quantitive calculations of
wave-particle interactions in the outer heliosphere by \citet{chalov06b} or \citet{fahr07} in which a consistent treatment of pick-up
ion transport in phasespace and wave turbulence transport in $k$-space was
presented. From these calculations it becomes evident that in fact in the
outer heliosphere pick-up ions drive turbulence powers, but are not
efficiently enough profitting from energy diffusion to produce power-law
tails.

The only way to fill up ion tails at energies larger than 1 keV resulting in
some power-law distribution, as it seems to us, is to inject ions into the
PUI regime at higher energies of about $\geq $50 keV and then let them cool
to smaller energies at their co-convection with the solar wind. This would
be a process similar to the one of pick-up ions originating from freshly
ionized neutral atoms which are injected at about 1 keV \citep[see][]{fahr07b,siewert08}. In the following, we study such a high-energy
injection mechanism to the PUI energy regime resulting from ACR ions that are adiabatically cooled to lower energies and, therefore, can be called ACR-PUIs.

\section{Injection to the PUI regime from the high-energy side}\label{sect4}

Here we first want to study the modulated part of the anomalous cosmic rays
(ACRs) with origin near the solar wind termination shock and investigate
whether they can perhaps serve as a possible injection seed from high
energies into the pick-up ion regime. We start from the analytic
expression for the modulated ACR spectral intensity given for the case of a
spherically symmetric solar modulation. The differential ACR intensity for
this case is given by the following expression \citep{stawicki00}
\begin{eqnarray}
j(r,p)&=&p^{2}f(r,p) \nonumber \\[0.5cm]
&=&\frac{3}{b}\int \mathrm dr_{0}\int \mathrm dp_{0}\frac{S(r_{0},p_{0})}{U}\frac{p_{0}y_{0}}{%
\phi }\left( \frac{r_{0}}{r}\right) ^{\frac{1+\delta }{2}}\left(\frac{p_{0}}{p}\right)^{%
\frac{3\delta -4\gamma -5}{2(2+\gamma )}}\times \nonumber \\
& \times & \exp \left(-\frac{y_{0}(1+h^{2})}{\phi }\right)
 I_{\frac{1+\delta }{1+\gamma -\delta }}(\frac{2y_{0}h}{\phi })
\end{eqnarray}
where $S(r_{0},p_{0})$ denotes the ACR source function with source
coordinates $r_{0}$ and $p_{0}$, $I_{\mathrm n}$ is the modified Bessel function of
the first kind, $U$ denotes the solar wind bulk velocity, and $h$ and $\phi $
have the following definitions
\begin{equation}
h=\left( \frac{r}{r_{0}}\right) ^{\frac{1+\gamma-\delta}{2}}\left(\frac{p}{p_{0}}%
\right)^{\frac{3}{2b}}
\end{equation}
\begin{equation}
\phi=1-\left(\frac{p}{p_{0}}\right)^{\frac{3\nu}{2+\gamma}}
\end{equation}
with the following additional quantities
\begin{equation}
y_{0}=\frac{\nu}{(1+\gamma-\delta)^{2}}\frac{r_{0}U}{\kappa_{rr,0}}
\end{equation}
\begin{equation}
b=\frac{2+\gamma}{(1+\gamma-\delta)}
\end{equation}
\begin{equation}
\nu=\left( 1+\gamma-\delta+\frac{2+\gamma}{3}\epsilon\right)
\end{equation}
and the following assumptions of the $r-$ and $p-$dependences:
\begin{equation}
U(r)\sim r^{\gamma},\,\, \kappa_{rr}(r,p)=\kappa_{rr,0} \left(\frac{r}{r_{\mathrm E}}\right)^{\delta} \left(\frac{p}{p_{\mathrm E}}\right)^{\epsilon}
\end{equation}
with an arbitrary reference momentum $p_{\mathrm R}$ for the bulk velocity and the spatial diffusion coefficient $\kappa_{rr}$.

As shown by \citet{stawicki00}, the above expression can be simplified
for low values of ACR particle momenta $p$ and then yields the following
expression for the ACR distribution
\begin{equation}
f(r,p)\simeq \frac{3}{b\Gamma(b)}\int \mathrm dr_{0}\int \mathrm dp_{0}\frac
{S(r_{0},p_{0})}{p_{0}U}y_{0}^{b}\exp(-y_{0})
\end{equation}
The modulation of the ACR spectrum due to spatial diffusion is contained in $y_0$. This shows that, hence, the ansatz by \citet{stawicki00} is an analytical way to express the modulated distribution function of the ACRs in the heliosphere. In the frame of its accuracy, this approach yields a valuable way to take the modulated ACR spectrum as the source of our cooling mechanism.
The spectrum of the ACR particles at the solar wind
termination shock serves as the ACR source function and, thus with good reasons, one adopts the
function developed by \citet{drury83} for an ideally planar shock which is
given by
\begin{eqnarray}
S(r_{0},p_{0})&=&\frac{1}{r_{0}^{2}}\frac{\partial }{\partial r_{0}} r_{0}^{2}\zeta _{0}  \\
&=&\frac{1}{r_{0}^{2}}\frac{\partial }{\partial r_{0}} r_{0}^{2}\left[S_{\mathrm{ACR},0}\cdot \left(\frac{p_0}{p_{\mathrm i}}\right)^{-q}\cdot \exp \left(-\frac{p_0}{p_{\max }}\right)\delta
(r_{0}-r_{\mathrm{sh}})\right],\nonumber
\end{eqnarray}
where the spectral index $q$ is connected with the shock compression ratio $%
s $ by $q=3s/(s-1)$. $p_{\mathrm i}$, $p_{\max }$ are appropriate values of the
Fermi-1 injection momentum and of the upper cut-off momentum, respectively.
This then leads to
\begin{eqnarray}
f(r,p) &\simeq&  \frac{3 S_{\mathrm{ACR},0}}{b\Gamma (b)U}%
\int \limits_{0}^{\infty}\frac{\mathrm dp_{0}}{p_{\mathrm i}}\left(\frac{p_{0}}{p_{\mathrm i}}\right)^{-q-1} \exp \left(-\frac{%
p_{0}}{p_{\max }}\right)\times \nonumber \\
&\times &\left(\frac{\nu }{(1+\gamma -\delta )^{2}}\frac{r_{\mathrm{sh}}U}{\kappa _{rr,\mathrm{sh}}}%
\right)^{b}\exp \left(-\frac{\nu }{(1+\gamma -\delta )^{2}}\frac{r_{\mathrm{sh}}U}{\kappa
_{rr},\mathrm{sh}}\right)
\end{eqnarray}

This expressions represents a constant distribution function. Numerical simulations by \citet{fichtner99} show a linear dependence of the ACR flux on the energy ($j_E\sim E$). This means that the distribution function in phase-space is constant which can be shown by considerations about normalization of the differential flux and the distribution function:
\begin{equation}\label{fconst}
f=4 \pi \frac{m j_E}{p^3}\frac{\mathrm dE}{\mathrm dp}\sim \frac{j_E}{E}=\mathrm{const.}
\end{equation}

Therefore, the $f(r,p)$ of the ACR particles is given by a constant value which follows from modulation theory:
\begin{equation}
f(r,p) \simeq f_{\mathrm{const}}.
\end{equation}

From the above expression we develop the total streaming $S(r,p_{\mathrm{pui},\mathrm i})$  \citep[see][]{gleeson67,fahr90,fahr_verscharen} for the ACR-PUIs in the range between the injection momentum $p_{\mathrm i}$ and an upper border $p_{\mathrm{pui},\mathrm i}$.  ACR particles with higher momenta propagate only according to the transport equation for cosmic rays and do not participate in the diffusive injection. With these considerations we find
\begin{equation}\label{streaming}
S=\int \limits_{p_{\mathrm i}}^{p_{\mathrm{pui,i}}}p^2 U \left(\frac{1}{3}f-\frac{1}{3p}\frac{\partial}{\partial p} f \right)\mathrm dp
=\frac{\left(p_{\mathrm{pui,i}}^3-p_{\mathrm i}^3 \right)} {9} Uf_{\mathrm{const}},
\end{equation}
as the total particle streaming at each place $r$.
Since we expect that $p_{\mathrm{pui,i}}\gg p_{\mathrm i}$, the lower border $p_{\mathrm i}$ can be neglected in Eq.~(\ref{streaming}).

Now we consider the local spatial divergence of this ACR-PUI streaming at some
lower momentum border and take this to be the high-energy injection source $Q$
to the PUI regime. With this choice, we obtain
\begin{equation}\label{quelle}
Q=\frac{1}{r^2}\frac{\partial}{\partial r}r^2 S=\frac{2p_{\mathrm{pui,i}}^3 Uf_{\mathrm{const}} } { 9 r }
\end{equation}
as the divergence of the ACR particle flux, constituting a PUI source.

\section{PUI Boltzmann-Vlasov equation with ACR-induced injection}\label{sect5}

In the ``solar'' rest frame (SF), the representative Boltzmann-Vlasov equation (BVE) is given by
\begin{equation}
(\vec{U}\cdot \nabla _{r})f+\left(\left.\frac{\mathrm d\vec{v}}{\mathrm dt}\right|_{\mathrm m}\cdot \nabla _{v}\right)f=U%
\frac{\partial f}{\partial r}+U\left.\frac{\partial v}{\partial r}\right|_{\mathrm m}\frac{\partial f}{\partial v}=P(r,v).
\end{equation}
under the prevailing conditions in the outer heliosphere, i.e.~negligible energy diffusion, for the stationary case \citep{fahr07b}.
The corresponding BVE equation in the ``solar wind'' rest frame (WF) has the form \citep{fahr07b}
\begin{equation}
\frac{\partial\tilde{f}}{\partial t}+\frac{1}{v^{2}}\frac{\partial}{\partial
v}\left(v^{2}\dot{v}_{\mathrm m}\tilde{f}\right)=\tilde{P}(t,v)
\end{equation}
where the second term on the left hand side describes the velocity-space
divergence of the phasespace flow connected with the magnetically induced
deceleration (i.e.~magnetic cooling), which is indicated by the subscript $\mathrm m$. The coordinate $t=t(r)$ denotes the proper time in the co-moving reference frame (WF).
Furthermore, $\tilde{P}(r,v)$ is the local ion injection rate to the PUI
regime.

For low energies ($\leq$ 1 keV), this rate is due to locally freshly ionized
neutral H-atoms and is given by $\tilde{P}(r,v)=\beta (r)\frac{1}{4\pi v^{2}U%
}\delta (v-U)$ with $\beta (r)$ being the local pick-up ion production rate \citep{fahr_siewert08}.

For high energies ($\geq$ 50 keV for the ACR-PUIs), the relevant injection, in contrast to the
normal sub-keV PUIs, in our view is given by the upper expression derived
from the modulated ACR spectrum and given by Eq.~(\ref{quelle}).

Furthermore, $\dot{v}_{\mathrm m}$ is determined by the magnetically induced velocity
decrease of particles with a velocity $v$, when they are convected outwards with
the solar wind bulk flow at a mean velocity $U$ to larger distances where
the co-convected interplanetary magnetic field $B$ appears reduced in
magnitude \citep{fahr07b}. At larger distances $r\geq$ 5 AU near the ecliptic,
the magnetic field decreases like $(1/r)$ (i.e.~in case of the nearly
azimuthal, distant Parker field). Under these conditions, one finds the
following $v$-dependent magnetic velocity-space drift
\begin{equation}
\dot{v}_{\mathrm m}=-U\frac{v}{r}
\end{equation}
and its associated radial gradient
\begin{equation}
\frac{\partial v_{\mathrm m}}{\partial r}=\frac{1}{U}\dot{v}_{\mathrm m}=-\frac{v}{r}
\end{equation}
in the form given by \citet{fahr_siewert08}.

Ions which are picked up at $r_{v}$ with a velocity $U$  will, without
other processes being involved, have ``magnetically'' cooled down to a
velocity $v$ at $r$ if the relation $r_{v}(v)=rv/U$ is fulfilled. The
injection of freshly created pick-up ions at $r_{v}(v)$ with an initial
velocity $v=U$ will be responsible for ions with velocity $v$ at $r$.

Taking all these constraints together, one finally finds, when reminding that
the time and distance coordinates are related to eachother by $\mathrm dr=U\mathrm dt$, that
for velocities $v\leq U$ the solution for $\tilde{f}$  in the WF is given
by \citep[see][]{fahr07b,siewert08}
\begin{equation}\label{fle}
\tilde{f}_{\leq }=\frac{1}{2\pi }\frac{r\beta \left(\frac{v}{U}r\right)}{U}v^{-3}.
\end{equation}

This distribution function $\tilde{f}_{\leq }$ can be evaluated for larger
solar distances $r\geq r_{0}=5\,\mathrm{AU}$ in the upwind hemisphere for near-ecliptic
positions assuming that at such solar distances the upwind H-atom density
can be considered as essentially constant, meaning that $n_{\mathrm H}\left(\frac{v}{U}r\right)=n_{\mathrm H}(r)=n_{\mathrm H,\infty }$ . This in fact is an acceptable approximation for
solar distances $r\geq$ 5 AU and velocities $1U\geq v\geq 0.2U$, and then
leads to
\begin{equation}
\tilde{f}_{\leq }=\frac{r}{2\pi U}\nu _{\mathrm{ex,E}}r_{\mathrm E}^{2}\left(\frac{v}{U}%
r\right)^{-2}n_{\mathrm H,\infty }v^{-3}=\frac{\nu _{\mathrm{ex,E}}r_{\mathrm E}^{2}U}{2\pi r}n_{\mathrm H,\infty
}v^{-5}
\end{equation}
i.e.~to the astonishing fact that under pure magnetic cooling, the resulting
PUI distribution function for velocities $v\leq 1U$ is a power law with the
interesting power index $\alpha =-5$ predicted and confirmed by \citet{fisk06,fisk07}; however, in their case expected as result of an
assumed quasi-equilibrium state established between magnetoacoustically
driven ion energy diffusion and magnetoacoustic turbulence generation.

The above result is valid only for ions with $v\leq U$. One possibility that
we now start to see here is that for ions with velocities $v\geq 1U$, i.e.~much higher than the original PUI injection threshold, one has to consider in addition some high-energy injection rate due to modulated ACRs as we
have derived above.

For those ACR-induced ions (i.e.~for the ACR-PUIs) with the relevant source function
\begin{equation}
Q=\frac{2p_{\mathrm{pui,i}}^3 Uf_{\mathrm{const}}} { 9 r },
\end{equation}
we find analogously to Eq.~(\ref{fle}) for ACR-PUIs
\begin{equation}
\tilde{f}_{\geq }=\frac{1}{2\pi }\frac{rQ\left(r\frac{v}{v_{\mathrm{pui},\mathrm i}}\right)%
}{U}v^{-3}=\frac{1}{2\pi }\frac{r\cdot \left[ \frac{2p_{\mathrm{pui,i}}^3 \frac{v_{\mathrm{pui},\mathrm i}} {v}} { 9 r } Uf_{\mathrm{const}} \right]}{U}v^{-3},
\end{equation}
which finally leads to an $r$-independent distribution function for the range $%
v\geq U$ in the form
\begin{equation}
\tilde{f}_{\geq }=\frac{1}{9\pi }m_{\mathrm p}^3 v_{\mathrm{pui},\mathrm i}^4 f_{\mathrm{const}} v^{-4}
\end{equation}

Alltogether, thus, we obtain the total PUI distribution in the following form:
\begin{eqnarray}
\tilde{f}&=&\tilde{f}_{\leq }+\tilde{f}_{\geq } \nonumber \\
&=&\frac{\nu _{\mathrm{ex,E}}r_{\mathrm E}^{2}U}{%
2\pi r}n_{\mathrm H,\infty }v^{-5}H(U-v)+ \frac{1}{9\pi } m_{\mathrm p}^3 v_{\mathrm{pui},\mathrm i}^4f_{\mathrm{const}} v^{-4}
\end{eqnarray}

where $H(x)$ is the well-known step function with $H(x\leq 0)=0$ and $%
H(x\geq 0)=1$.

In any case, the above result shows that a velocity power-index of ``$-5$'' is
obtained for the velocity-range $v\leq U\leq v_{\mathrm{pui},\mathrm i}$ and a power index of
``$-4$'' is obtained for the velocity-range $U\leq v\leq v_{\mathrm{pui},\mathrm i}$ where $%
v_{\mathrm{pui},\mathrm i}=p_{\mathrm{pui},\mathrm i}/m_{\mathrm p}$.

Rescaling velocity in units of $U$ (i.e.~$X=v/U$), leads to the total distribution function
\begin{eqnarray}
\tilde{f}&=&\frac{\left(\sigma _{\mathrm{ex}}n_{\mathrm{sE}}U\right)r_{\mathrm E}^{2}}{2\pi rU^{4}}%
n_{\mathrm H,\infty }X^{-5}H(1-X)+\frac{1}{9\pi }m_{\mathrm p}^3 X_{\mathrm{pui},\mathrm i}^4f_{\mathrm{const}}X^{-4} \nonumber \\
&=&\frac{n_{\mathrm{sE}}}{2\pi U^{3}}\left[\Lambda \left(\frac{r_{\mathrm E}}{r}\right)X^{-5}H(1-X)+
\frac{2}{9\pi } \frac{m_{\mathrm p}^3 X_{\mathrm{pui},\mathrm i}^4  U^3 f_{\mathrm{const}}}{n_{\mathrm{s,E}}}X^{-4}
\right]
\end{eqnarray}

We introduce now two dimensionless characterizing quantities. First, we define
\begin{equation}
\Lambda =(\sigma _{\mathrm{ex}}n_{\mathrm H,\infty }r_{\mathrm E})\simeq 2.3\times 10^{-3},
\end{equation}
where the charge exchange cross section $\sigma _{\mathrm{ex}}\simeq 10^{-15}\,\mathrm{cm}^2$ \citep{fahr71} and the hydrogen density $n_{\mathrm H,\infty }\simeq 0.15 \,\mathrm{cm}^{-3}$ \citep{izmodenov03} are used.

With 
\begin{equation}
\Delta =\frac{2}{9\pi } \frac{m_{\mathrm p}^3 X_{\mathrm{pui},\mathrm i}^4 U^3 f_{\mathrm{const}}}{n_{\mathrm{s,E}}},
\end{equation}
one can finally find
\begin{equation}\label{spektrum}
\tilde{f}(r,v)=\frac{n_{\mathrm{sE}}}{2\pi U^{3}}\left[\Lambda \left(\frac{r_{\mathrm E}}{r}\right)\cdot
X^{-5}H(1-X)+\Delta \cdot X^{-4}\right].
\end{equation}

Measurements by the SWICS instrument on the ULYSSES space probe show a PUI phase space density of about $100\,\mathrm s^3\,\mathrm{km}^{-6}$ at two times the solar wind velocity at a distance of $5.26$ AU \citep{gloeckler03}. Eq.~(\ref{spektrum}) leads to a value for the PUI phase space density of about $\tilde f_{\leq}\simeq 155\,\mathrm s^3\,\mathrm{km}^{-6}$, which is in good accordance to the observations. The difference could be a consequence of the assumption of a constant $n_{\mathrm H}$, especially at small distances from the sun.

\begin{figure}[t]
\vspace*{2mm}
\par
\begin{center}
\includegraphics[angle=-90,width=\columnwidth]{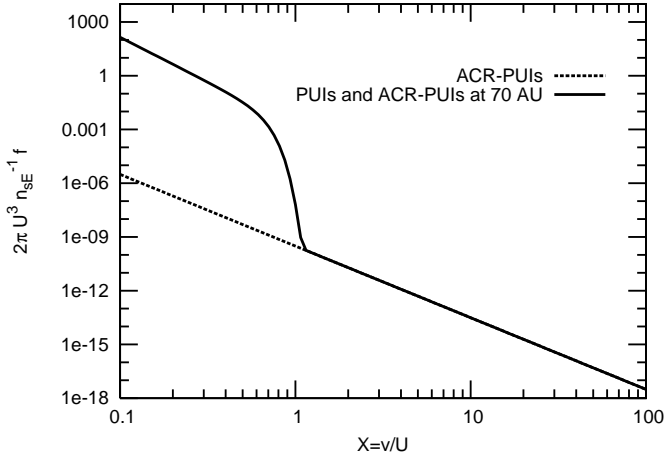}
\end{center}
\caption{Phase space density spectra for PUIs and ACR-PUIs. The PUI distribution is dependent on the distance from sun, whereas the ACR-PUIs are not. PUIs cannot gain velocities above $U$, i.e.~$X=1$. The solid line is led down to the ACR-PUI spectrum at the cut-off at $X=1$ artificially.}
\label{fig_spectra}
\end{figure}

Observations by the LECP instruments on VOYAGER 1 and 2 \citep{lanzerotti01} show an $r$-independent proton intensity in the energy range 0.57-1.78 MeV of about $j_E\simeq 4\times 10^{-4}$ particles per (cm$^2$ s sr MeV), which corresponds to $j_E\simeq 250$ particles per (cm$^2$ s sr erg) in Gaussian units. This value can be converted to a phase space density (cf.~Eq.~(\ref{fconst})):
\begin{equation}
\tilde f_{\geq}=4\pi \frac{m_{\mathrm p}^3j_E}{p^2(1\,\mathrm{MeV})}=2.7\times 10^{-39} \,\mathrm s^3\, \mathrm{cm}^{-6}
\end{equation}
If we assume that these ions observed by \citet{lanzerotti01} are ACR-PUI particles with the proposed behaviour, this leads to a value for $\Delta$ obtained from the measurements of
\begin{equation}
\Delta=\frac{2\pi U^3}{n_{\mathrm{sE}}}X^4(1\,\mathrm{MeV})\tilde f_{\geq}=3.1\times 10^{-10}.
\end{equation}
We take this observational value to avoid uncertainties at the determination of the ACR intensity $f_{\mathrm{const}}$ and the injection border $X_{\mathrm{pui},\mathrm i}$. The calculated spectra are shown in Fig.~\ref{fig_spectra}.

\section{Comparison of our results with data}\label{sect6}

Now we want to compare the above results with Voyager and Ulysses spectral
ion data. For that purpose, we first transform the above velocity
distribution function into a distribution of spectral energy flux $j(E)$
given in units of [ions/(cm$^{2}$ s sr MeV)] . This then leads to
\begin{eqnarray}
j(E)&\sim & \sqrt{E}\cdot f(E)\sim \sqrt{E}\left(\frac{E}{E_{0}}\right)^{(1-\alpha
_{v})/2} \nonumber \\
&\sim &\left(\frac{E}{E_{0}}\right)^{(2-\alpha _{v})/2}\sim \left(\frac{E}{E_{0}}\right)^{-\alpha _{E}}
\end{eqnarray}
where $\alpha _{v}$ and $\alpha _{E}$ are the velocity- and the energy-power indices, respectively. As one can easily see, $\alpha _{v}=5$ leads to $\alpha
_{E}=1.5$, whereas $\alpha _{v}=4$ would lead to $\alpha _{E}=1.0$. Looking
into the data obtained by VOYAGER 1 during the year 2004 \citep[see][]{decker05,decker06} before the shock crossing occured, one can find -- even though the data are very time-variable in this intervall -- that for ions with higher energy
$(E\simeq 10^{3}\,\mathrm{keV})$ the energy power index has been observed with $%
\alpha _{E}(E\geq 200\,\mathrm{keV})\simeq 1.5$, whereas interestingly enough for ions with lower
energy (i.e.~$E\simeq 40\,\mathrm{keV})$ a smaller power index of $\alpha
_{E}(E\leq 200\,\mathrm{keV})\simeq 1.0$ seems to be indicated. That supports
our above derived theoretical prediction that the velocity power index of
ACR-induced pick-up ions in the outer heliosphere is $\alpha _{v}\simeq 4$,
rather than $\alpha _{v}\simeq 5$ as for the ions with higher energy (MeV).

Also during the most recent VOYAGER-2 crossing of the shock \citep[see][]{decker08} it became evident that the energy-power indices registered both before and after the shock crossing show values of $\alpha _{E,1}\simeq
\alpha _{E,2}\simeq 1.2\pm 0.2$ which also nicely confirms a velocity index
of $\alpha _{v}\simeq 4$.

In addition, it is important to recognize that the particle detectors of the VOYAGER spacecraft did not see radial
changes of the spectral flux intensity $j(E)$ at the lower energy channels (0.35-1.5 MeV) during the period from years 1995 through 2005 \citep{lanzerotti01,krimigis03,decker03}. This is also a support for the newly derived
theoretical expression given in Eq.~(\ref{spektrum}) and showing that a
distance-independent intensity can be expected at energies $5\,\mathrm{keV}\leq E\leq
100$ keV. Only the part of the spectrum that is a result of cooled pick-up ions shows a dependence on the distance due to the different PUI abundances. The ACR-PUIs that are injected from ACRs and cooled from high energies, however, are proven to have no radial variation.

Our model explains occuring power-law tails with an index $-4$. There are, however, observations around 5 AU undoubtedly showing a power-index $-5$ for suprathermal ion tails \citep[e.g.,][]{gloeckler03}. As we can show, the favored explanation by \citet{fisk06,fisk07} cannot hold in the described way. Maybe other processes (such as anomalous Fermi-2 type wave-particle energy diffusion), which are not treated in our approach, can lead to a power-index of $-5$ at energies just above 5 keV. But this interaction does not explain the occurence of extended power-law tails at larger distances \citep{chalov04}.

It is perhaps still a little bit an open question, where, i.e.~at what
lower energy, one should cut off the modulated ACR spectrum to calculate the
injection to the PUI regime. Even though this does not count very much in
quantitative terms, since the lower energy part of the modulated ACR
spectrum has in fact a constant distribution function (see Eq.~(\ref{fconst})), it may nevertheless
represent an intellectually interesting question, to decide up to what
energies convection and Fermi-2 energy-diffusion are dominant, and from what
energies upwards Fermi-1 acceleration and spatial diffusion dominate. There
is one clear hint given to answer this question, namely by the
momentum-dependence of the spatial diffusion coefficient $\kappa_{rr} (r,p)$
generally given in the form
\begin{equation}
\kappa (r,p)=\kappa _{0}\kappa _{r}(r)\kappa _{p}(p)\sim \left(\frac{r}{r_{\mathrm E}}\right)^{\delta
}\left(\frac{p}{p_{\mathrm R}}\right)^{\epsilon}
\end{equation}
where, dependent on the momentum-range, the exponent $\gamma $ is expected
to be in the range $1\leq \gamma \leq 2$ \citep[see][]{leroux92,jokipii96}. This shows that spatial diffusion becomes less and less
efficient, the lower the ion momentum $p$ is. Below some critical value $%
p_{\mathrm c}$, ions loose their degree of kinetic freedom to spatially diffuse
relative to the solar wind background flow and, thus, they are simply 
convected outwards with the solar wind bulk flow then. 

The determination of the absolute height of the ACR-PUI spectrum is a slightly problematic endeavor. In our calculation, we normalize the absolute spectrum with the aid of observational data. The ACR intensity $f_{\mathrm{const}}$ and the limit for the momentum, up to which the injection is effective due to the total streaming (see Eq.~(\ref{streaming})), are quite uncertain. Another open question results from a lack of data. Especially, the transition from the PUI to the ACR-PUI part has not been covered sufficiently by particle detectors in the outer parts of the heliosphere. Maybe future missions can provide a closer look to this part of the spectrum.

\begin{acknowledgements}
      H.-J.~F.~and D.~V.~are grateful to the
      Deut\-sche For\-schungs\-ge\-mein\-schaft for financial support within the frame
      of the DFG project Fa~97/31-2. One of us, I.~V.~C., is grateful for financial support in the frame of DFG project 436 RUS113/110/0-4.
\end{acknowledgements}

\bibliographystyle{aa}
\bibliography{10755}

\end{document}